\documentclass[twocolumn]{article}
\pdfoutput=1

\usepackage{hyperref}
\usepackage{abstract}
\usepackage{amsmath}
\usepackage{amsmath,upgreek}
\usepackage{amssymb}
\usepackage{authblk}
\usepackage[backend=biber,style=phys,autocite=plain,sorting=none]{biblatex}
\usepackage{bm}
\usepackage{caption}
\usepackage{graphicx}
\usepackage{makecell}
\usepackage{mathtools}
\usepackage{multicol}
\usepackage{physics}
\usepackage{siunitx}
\usepackage[caption=false]{subfig}
\usepackage{textcomp}

\bibliography{References} 
\graphicspath{{'../Figures/'}}

\newcommand\figdc[2]{
	\begin{figure*}
		\centering
		\includegraphics[width=17.1cm]{Figures/#1}
		\caption{#2} 
		\label{fgr:#1}
	\end{figure*}
}

\hypersetup{
	colorlinks=true,
	linkcolor=black,
	citecolor=black,
	filecolor=black,
	urlcolor=black,
}

\DeclarePairedDelimiter\floor{\lfloor}{\rfloor}

\date{}

\title{Mesoscale modelling of polymer aggregate digestion}

\author[1,2]{Javor K. Novev\thanks{yavor.novev@nbi.ku.dk}}
\author[1,2]{Amin Doostmohammadi}
\author[3]{Andreas Z\"{o}ttl}
\author[1]{Julia M. Yeomans}
\affil[1]{The Rudolf Peierls Centre for Theoretical Physics, University of Oxford, Oxford, OX1 3PU, UK}
\affil[2]{Present address:   Niels Bohr Institute, University of Copenhagen, Blegdamsvej 17, 2100 K\o benhavn \O, Denmark}
\affil[3]{Institute for Theoretical Physics, TU Wien, Wiedner Hauptstraße 8-10, Wien, Austria}
\begin{document}

\twocolumn[
	\maketitle
	\begin{onecolabstract}
		We use mesoscale simulations to gain insight into the digestion of biopolymers by studying the break-up dynamics of polymer aggregates (boluses) bound by physical cross-links.  
		We  investigate aggregate evolution, establishing that the linking bead fraction and the interaction energy are the main parameters controlling stability with respect to diffusion. We show \textit{via} a simplified model that chemical breakdown of the constituent molecules causes aggregates that would otherwise be stable to disperse. We further investigate breakdown of biopolymer aggregates in the presence of fluid flow. Shear flow in the absence of chemical breakdown induces three different regimes depending on the flow Weissenberg number ($Wi$). i) At $Wi \ll 1$, shear flow has a negligible effect on the aggregates. ii) At $Wi \sim 1$, the aggregates behave approximately as solid bodies and move and rotate with the flow. iii) At $Wi \gg 1$, the energy input due to shear overcomes the attractive cross-linking interactions and the boluses are broken up. Finally, we study bolus evolution under the combined action of shear flow and chemical breakdown, demonstrating a synergistic effect between the two at high reaction rates.
	\end{onecolabstract}
]

\saythanks

\section{Introduction}
	Food digestion is a complex cascade of chemical and physical processes spanning multiple length scales \cite{Bornhorst2016, Wang2010}, ranging from centimetric dimensions for food ingested through the mouth down to the molecular scale for nutrients broken down and absorbed in the stomach and the intestines. As studies have established links between dietary habits and many common health issues, e.g., obesity and diabetes, \cite{Bornhorst2016}, understanding digestion is of broad biomedical relevance. Specifically, a better insight into the process would facilitate the design of so-called `functional foods' that aside from nutritional benefits are associated with an improved state of health and/or reduction of the risk of some diseases \cite{Viuda-Martos2010, Ozen2012}. The effect of functional and other foods on health is to a large extent controlled by their rate of digestion;  for example, resistant starches, i.e., ones that are not broken down until they are transported to the large intestine, are known to have beneficial health effects \cite{Bird2009}. 

	Food digestion starts with mastication in the mouth. There, food is broken down mechanically, lubricated by saliva and converted into a cohesive mass known as a bolus \cite{Singh2015, Bornhorst2012} that consists of particles of typical size of $\sim \SI{1}{\milli\meter}$ \cite{Jalabert-Malbos2007, Bornhorst2012}. The bolus is then transported to the stomach where it is broken down chemically through hydrolysis by the hydrochloric acid and the enzymes in the gastric juices, as well as mechanically \textit{via} muscular contractions; limited nutrient adsorption also occurs in the stomach \cite{Singh2015}. After gastric sieving, which only allows particles of size smaller than $\sim \SI{1}{\milli\meter}$ to pass \cite{Schulze2006}, the mixture of partially broken down food and secretions from the digestive tract (chyme) is then transported to the small intestine where the pH is neutral to basic \cite{Boland2016}. The small intestine is the site of further enzymatic and mechanical breakdown and much of the absorption of nutrients occurs there \cite{Cleary2015}. Finally, a low-viscosity slurry reaches the large intestine where the key processes are microbial breakdown and water absorption \cite{Cleary2015}. \par	

	Given the complicated and multiscale nature of the digestive pathway, and the varied structure of the biopolymeric molecules relevant to the human diet, formulating models of digestion is very demanding and progress is only likely to be made by using a range of methods relevant to different length scales. Much of the existing work is based on continuum  approaches, see e.g. \cite{Taghipoor2012,Moxon2016} and the reviews in Refs. \cite{Cleary2015, Ferrua2011, Ferrua2015}.   In particular, some authors have formulated anatomically accurate 3D computational models of food breakdown in the mouth \cite{Cleary2015} and of gastric digestion \cite{Cleary2015, Ferrua2010, Ferrua2011, Ferrua2015}. Other authors have also worked towards developing {\it in vitro} models for the experimental study of food digestion, e.g. the gastric simulator described in Refs. \cite{Kong2008, Kong2010}. Although some effort has been dedicated to studying digestion at the mesoscale, for example \textit{via} the lattice Boltzmann method \cite{Wang2010}, to the best of our knowledge, the approach described of this article, using a mesoscale algorithm to study the dynamics and stability of the polymeric aggregates that arise in the intermediate stages of the breakdown of complex carbohydrates, has not yet been explored. The method we use explicitly models the flow of polymer-solvent mixtures, eliminating the need for approximate constitutive relations and the corresponding assumptions. However, feasible simulations do require coarse-graining molecular detail. \par		

	We focus on the dynamics of polymeric aggregates in a viscous medium. Although we do not aim to mimic physiological conditions exactly, our results are most pertinent to food digestion in the stomach. Rather than attempting to simulate a specific biopolymer or to describe the complex mixture of biomolecules encountered in the digestive tract, we use a simple model that allows insight into some of the generic mechanisms controlling digestion. Our polymeric aggregates, which we refer to as boluses, are initially approximately spherical in shape and consist of monodisperse linear bead-spring chains with no bending rigidity, and we assume that a fraction of the beads can form physical cross-links. (Note that we use the term `bolus' to refer to a generic aggregated mass rather than in its specific sense of a lubricated conglomerate of typically millimetric food particles formed during mastication \cite{Jalabert-Malbos2007, Bornhorst2012}).

	We first consider the dynamics of such boluses in a quiescent fluid, determining the key parameters that control whether they are stable with respect to diffusion. We next study how two of the major factors at play in the digestive tract, chemical breakdown of the polymers and shear flow, affect bolus evolution. Finally, we investigate the combined action of simple shear and chemical breakdown of the polymers.  The predictions of our model, while not directly comparable to a specific system, are relevant to the digestion of starch, which is a major energy source in the typical human diet \cite{Shrestha2015, Liu2017}. \par

\section{Methods} \label{Methods}

	\subsection{Polymer Model}

		We consider a coarse-grained, bead-spring model of monodisperse linear polymer chains in which the bonds are approximated by harmonic springs, yielding the potential~\cite{Allen1987}:
		\begin{equation}\label{eq1}
			u_{\text{bond}} = \frac{1}{2}k_{\text{bond}}\sum_{\text{i}=2}^{N_{\text{bead}}}{\left(\abs{\Delta\bm{r}_{\text{i}}}-l_0 	\right)^2},
		\end{equation}
		with $\Delta\bm{r}_{\text{i}} = \bm{r}_{\text{i}} - \bm{r}_{\text{i}-1}$, where $\bm{r}_{\text{i}}$ is the position vector of bead $\text{i}$, $N_{\text{bead}}$ is the number of beads in a single chain, the resting bond length is $l_0 = \sigma$, $\sigma$ being the bead diameter, $k_{\text{bond}} = 10^5\textrm{k}_\textrm{B}T$, where $\textrm{k}_\textrm{B}$ is the Boltzmann constant and $T$ is the temperature.

		We model the inter- and intra-chain interactions between beads with the truncated Lennard-Jones potential~\cite{Allen1987}:
		\begin{equation}\label{eq3}
			u_{\text{LJ}} =
			\begin{cases}
				\epsilon + 4\epsilon\left[\left(\frac{\sigma^*}{r}\right)^{12}-\left(\frac{\sigma^*}{r}\right)^6\right] & r \leq r_{\text{cut-off}} \\
					0 & r > r_{\text{cut-off}} \\
			\end{cases}
		\end{equation}
		where $r$ is the distance between the centres of the beads and $\epsilon$ characterizes the interaction strength. For beads that do not form cross-links, we use the Weeks-Chandler-Andersen potential~\cite{Weeks1971}, which we obtain from Eq.~\eqref{eq3} by setting the cut-off radius to $\sigma$ and $\sigma^*$ to $\sigma/2^{1/6}$. This choice of interaction potential implies that we are modelling a good solvent~\cite{Ryder2006}. \par

		To model physical cross-linking, which can act both within and between polymer chains, we modify the Lennard-Jones potential so that interactions between individual polymer beads can be attractive. Upon generating the initial conditions for each simulation, there is a probability $p_{\text{link}}$ that each bead will be able to form links.  We model the force between two linking beads with an offset Lennard-Jones potential,
		\begin{equation}\label{eq4}
			u_{\text{link}}=u_{\text{LJ}}+u_{\text{offset}},
		\end{equation}
		where we choose the offset so that the force acting between the two beads, $\bm{F}=-\nabla u_{\text{link}}$, vanishes at $r = r_{\text{cut-off}}$,
		\begin{equation}\label{eq5}
			u_{\text{offset}} = 
				\begin{cases}
					-\frac{r^2}{2r_{\text{cut-off}}}\times\lim_{r\to r_{\text{cut-off}}^-}\frac{\partial u_{\text{LJ}}}{\partial r} & r \leq r_{\text{cut-off}} \\
					0 & r > r_{\text{cut-off}}. \\
				\end{cases}
		\end{equation}
		For linking beads, $\sigma^*=\sigma$ and $r_{\text{cut-off}}=(5/2)\sigma$ in both $u_{\text{LJ}}$ and 		$u_{\text{offset}}$. $u_{\text{link}}$ has a minimum of $\approx 10^{-2} \epsilon$ at approximately the same $r$ as $u_{\text{LJ}}$, $r_{\text{min}}\approx 2^{(1/6)}$. Offsetting the interaction potential in this way is standard in MD simulations, and a discussion of using an offset linear in $r$ can be found in Section 5.2.4 in Ref.~\cite{Allen1987}. Note that the physical cross-links, which act both within and between polymer molecules, are much weaker than the intramolecular chemical bonds holding the chains together. For comparison, if we take the Taylor series of $u_{\text{link}}$ about $r_{\text{min}}$ and truncate it to second order, we obtain an effective spring constant of $k_{\text{eff}} = u''(r_{\text{min}}) \approx 56.8 \epsilon$, with $\epsilon\sim\textrm{k}_\textrm{B}T$, whereas  $k_{\text{bond}} = 10^5\textrm{k}_\textrm{B}T$. \par

		In order to prevent the repulsive forces from diverging during initialization we normalize the forces arising from the WCA potential and $u_{\text{link}}$ in the following way: rewriting the force as $\bm{F}=-\nabla u = \bm{r}F_0$, with $\bm{r}$ being the vector connecting the centres of the two beads, for $\abs{F_0}>F_{\text{max}}=10^3$, we use the expression $\bm{F}=\bm{r}F_{\text{max}}F_0/\abs{F_0}$. The force normalization is only relevant if two beads overlap significantly. Our algorithm for bead initialization allows such overlaps, but they are quickly eliminated as the aggregates evolve and unlikely to occur elsewhere in our simulations as the bead-to-bead interaction potential becomes strongly repulsive at short distances. \par

		We generate the initial positions of the beads so that they form an aggregate (bolus) of approximately spherical shape with all beads within a sphere of radius $R_{\text{sphere}}=6\sigma$ whose centre coincides with that of the simulation box. We specify the desired polymer volume fraction $\rho$ in the sphere and generate $N_{\text{poly}}$ polymer chains such that $\rho$ is not exceeded, $N_{\text{poly}} = \floor*{6\rho V_{\text{sphere}}/\left(\uppi N_{\text{bead}}\sigma^3\right)}$, where the floor function $\floor*{x}$ acting on the real number $x$ returns the largest integer $\leq x$, see Ref. \cite{Hazewinkel2001}.  For each set of parameters, we perform $N_{\text{ens}} = 20$ simulations with different random initial conditions and average the results over the ensemble. \par

		The process of digestion involves enzymes that hydrolyse the chemical bonds between monomers and thus break down the polymeric chains. As a simple way of modelling the enzymatic hydrolysis, we introduce random bond cleavage to the model. We do this by introducing a rate constant $k_{\text{break}}$ that the bond between any two beads is broken at each streaming step; if a bond is cleaved, the two beads participating in it are no longer connected \textit{via} the harmonic potential \eqref{eq1}. All bonds are equally susceptible to attack: this mode of enzyme degradation of carbohydrates is known as a multichain attack in the literature \cite{Robyt1967, Bijttebier2008}.

	\subsection{Multi-particle Collision Dynamics}

		We simulate the flow \textit{via} multiparticle collision dynamics (MPCD), a mesoscale technique for solving the Navier-Stokes equations that treats the solvent as a collection of point particles which move ballistically during streaming steps and exchange momentum in collision steps~\cite{Yeomans2006}. The MPCD implementation used here follows Refs.~\cite{Zottl2019, Zottl2019a}, see references therein, in particular the work of Gompper et al. \cite{Gompper2009}. The main difference with the approach from Ref. \cite{Zottl2019, Zottl2019a} is that here we also include physical cross-linking. In a streaming step of duration $\updelta t$, a fluid particle i changes its position $\bm{r}_{\text{i}}$ according to	
		\begin{equation}\label{eq6}
			\bm{r}_{\text{i}}(t+\updelta t)=\bm{r}_{\text{i}}(t)+\bm{v}_{\text{i}}(t)\updelta t,		
		\end{equation}
		where $\bm{v}_{\text{i}}$ is the particle velocity.
		Between streaming steps, the particles are sorted in cubic cells of length $L_0 = \sigma$ and exchange momentum in collision steps according a collision rule that employs an Andersen thermostat~\cite{Zottl2014}:
		\begin{equation}\label{eq7}
			\bm{v}_{\text{i}}(t+\updelta t)=\bm{v}_{\text{cell}}(t)+\bm{v}_{\text{rand}}(t)+\bm{v}_{\text{P}}(t)+\bm{v}_{\text{L}}(t),
		\end{equation}
		where $\bm{v}_{\text{cell}}$ is the centre-of-mass velocity of the cell, $\bm{v}_{\text{rand}}$ is a random velocity obeying the Maxwell-Boltzmann distribution at temperature $T$, and the terms $\bm{v}_{\text{P}}$ and $\bm{v}_{\text{L}}$  ensure that linear and angular momentum are conserved~\cite{Zottl2019}. The parameters for the MPCD fluid are the same as in Ref.~\cite{Zottl2019} and correspond to a viscous flow with a low Reynolds number: the number density of the fluid particles is $\rho_{\text{MPCD}}=10/\sigma^3$, the time step is $\updelta t=0.02\sqrt{m\sigma^2/(\mathrm{k_{B}}T)}$, $m$ being the mass of an individual particle. The quantity $\tau_0 = \sqrt{m\sigma^2/(\mathrm{k_{B}}T)}$ has the dimension of time, and throughout the text, it is the implied time unit wherever no other is specified. The implied units of length, mass and energy are respectively $\sigma$, $m$ and $\mathrm{k_{B}}T$. \par

		During the streaming step, the forces acting on the polymer beads are calculated from Eqs.~\eqref{eq1}-\eqref{eq4} and their positions are computed \textit{via} the velocity Verlet algorithm~\cite{Allen1987} at intervals of $\updelta t_\text{poly}=\updelta t/50$. The polymer beads have mass $m_{\text{B}} = 10m$ and are included in the collision step~\cite{Zottl2019,Yeomans2006}. \par

		We study the temporal evolution of polymeric aggregates in an unbounded, initially quiescent fluid and in shear flow. To this end, we employ a cubic simulation box of size $L = 48\sigma$ and simulate a time period of $t_{\text{sim}}= 10^5\updelta t$. For simulations involving an unbounded fluid, we apply periodic boundary conditions to all walls. To simulate shear flow, we introduce planar solid walls that are situated at $x = \pm L/2$ and move with a velocity $v_z = \pm u_{\text{wall}}$. This corresponds to an approximately linear $v_z(x)$ with a shear rate $\dot{\gamma} = 2u_{\text{wall}}/L$.
		In simulations that involve solid walls, the walls contain virtual particles that interact with those of the fluid and polymer according to a bounce-back rule \cite{Zottl2014}. 

\section{Results}

	In the simulations we discuss below, we study the evolution of aggregates comprised of fully flexible polymer chains with a Kuhn length of $\sim \sigma$ comprised of  $N_{\text{bead}} = 20$ coarse-grained beads. We explore the effect of various parameters, namely, the fraction of beads that can form links ($p_{\text{link}}$), the interaction energy parameter ($\epsilon$), the rate of enzymatic hydrolysis ($k_{\text{break}}$) and the shear rate of imposed shear flows ($\dot{\gamma}$). We summarize the key parameters in Table~\ref{tbl:main_parameters}.\par

	\begin{table*}[t]
		\small
		\caption{\ Key parameters influencing bolus dynamics and respective ranges explored in this work}
		\label{tbl:main_parameters}
		\begin{tabular*}{\textwidth}{@{\extracolsep{\fill}}lll}
			\hline
			\makecell[l]{Parameter} & \makecell[l]{Significance} & \makecell[l]{Studied \\range} \\
			\hline
			\makecell[l]{$p_\mathrm{link}$} & \makecell[l]{probability that every individual bead \\ is initialized as a linking bead} & \makecell[l]{$0-1$} \\
			$\epsilon$ & interaction strength, see eqs.~\ref{eq3}-\ref{eq5} & $1-10$ \\
			\makecell[l]{$k_\mathrm{break}$} & \makecell[l]{probability that a bond \\ is broken in a given time step} & \makecell[l]{$0-10^{-3}$} \\
			\makecell[l]{$Wi$} & \makecell[l]{dimensionless shear rate, defined as \\ the product of the dimensional shear \\ rate and the bolus characteristic time, $\dot{\gamma}\langle t_{\text{bolus}}\rangle$} & \makecell[l]{$0-58$} \\
			\hline
		\end{tabular*}
	\end{table*}

	For all quantities of interest $X(t)$, we calculate the ensemble mean value,
	\begin{equation}\label{eq10}
		\langle X(t) \rangle = \frac{1}{N_{\text{ens}}}\sum_{\text{k}=1}^{N_{\text{ens}}}{X_{\text{k}} (t)},
	\end{equation}
	where k labels a given set of initial conditions.
	The position vector of the bolus centre of mass for the simulation with the set of initial conditions labelled k is 
	\begin{equation}\label{eq11}
		\bm{r}_{\text{av k}}(t) = \frac{1}{N_{\text{poly}}}\sum_{\text{i}=1}^{N_{\text{poly}}}{\bm{r}_{\text{CM ik}}(t)}.
	\end{equation}
	A useful quantity to characterize the shape and dimensions of the irregularly-shaped boluses is the gyration tensor. Its diagonal components are defined as
	\begin{equation}\label{eq12}
		G_{\text{ll k}}(t)=\frac{1}{N_{\text{tot}}}\sum_{\text{i}=1}^{N_{\text{poly}}}{\sum_{\text{j}=1}^{N_{\text{bead}}}{\left[\left(\bm{r}_{\text{ijk}}(t)-\bm{r}_{\text{av\ k}}(t)\right)\cdot\textbf{e}_{\text{l}}\right]^2}}, 
	\end{equation}
	where $N_{\text{tot}}=N_{\text{poly}}N_{\text{bead}}$ is the total number of beads in the bolus and $\textbf{e}_{\text{l}}$ is a Cartesian basis vector. Other authors have used analogous definitions of $G_{k}$ to characterize individual polymer molecules, see e.g. Mattice and Suter~\cite{Mattice1994} and Liebetreu et al.~\cite{Liebetreu2018}. In particular, gyration tensors have been defined for individual ring polymers \cite{Liebetreu2018, Chen2013} and single-chain polymeric nanoparticles \cite{Formanek2019}.\par
	The bolus gyration radius is related to the trace of the gyration tensor $G_\text{k}$ as
	\begin{equation}\label{eq13}
		R^2_{\text{g bol k}} = \mathrm{tr}\left(G_\text{k}(t)\right).
	\end{equation}
	For polymers that do not form cross-links, the individual molecules diffuse away and the aggregate disperses (Figure ~\ref{fgr:p_link_snapshots}), leading to  a $\langle R^2_{\mathrm{g \ bol}}(t) \rangle$ which is linear in $t$ at long times. In contrast, for boluses with a sufficiently high number of strong cross-links, $\langle R_{\mathrm{g \ bol}} \rangle$ approaches a stationary value. For this reason, the quantity $s = \partial \langle R^2_{\mathrm{g \ bol}}(t) \rangle/\partial t / \langle R^2_{\mathrm{g \ bol}}(0) \rangle$ which characterizes the rate of expansion of the bolus surface,  is useful in describing the different modes of bolus behaviour.\par

	Bolus evolution at short times is dominated by repulsive forces due to overlapping beads in the initial condition, leading to a sharp maximum of $\langle R^2_{\mathrm{g \ bol}} \rangle$  at $t\to 0$, see e.g. Figure~\ref{fgr:Rgbolsq_vs_t_diff_p_link_inset_dRgbolsqdt_vs_p_link}. This repulsive Lennard-Jones interaction is short-ranged and after going through this maximum in size, the bolus contracts due to attractive forces, typically going through a shallow minimum in $\langle R^2_{\mathrm{g \ bol}} \rangle$ (Figure~\ref{fgr:Rgbolsq_vs_t_diff_p_link_inset_dRgbolsqdt_vs_p_link}). At still longer time scales, at $t \gtrsim t_0 = t_\mathrm{sim}/20 = 100\tau_0$, $\langle R^2_{\mathrm{g \ bol}} \rangle$ is determined by the balance between cross-linking interactions holding the bolus together and the diffusion of the polymers driving the dispersion of the aggregate. As this is the regime we are interested in, we focus on $t > t_0$. \par  

	\subsection{Boluses in a quiescent fluid}

		We start by investigating the stability of boluses in a quiescent fluid and its dependence on the properties of the constituent polymers.

		\subsubsection{Varying the linking bead fraction $p_{\mathrm{link}}$~~}

\begin{figure*}
				\centering
				\subfloat[]
				{
					\label{fgr:snapshots_p_link_0}
					\includegraphics[width=.95\textwidth]{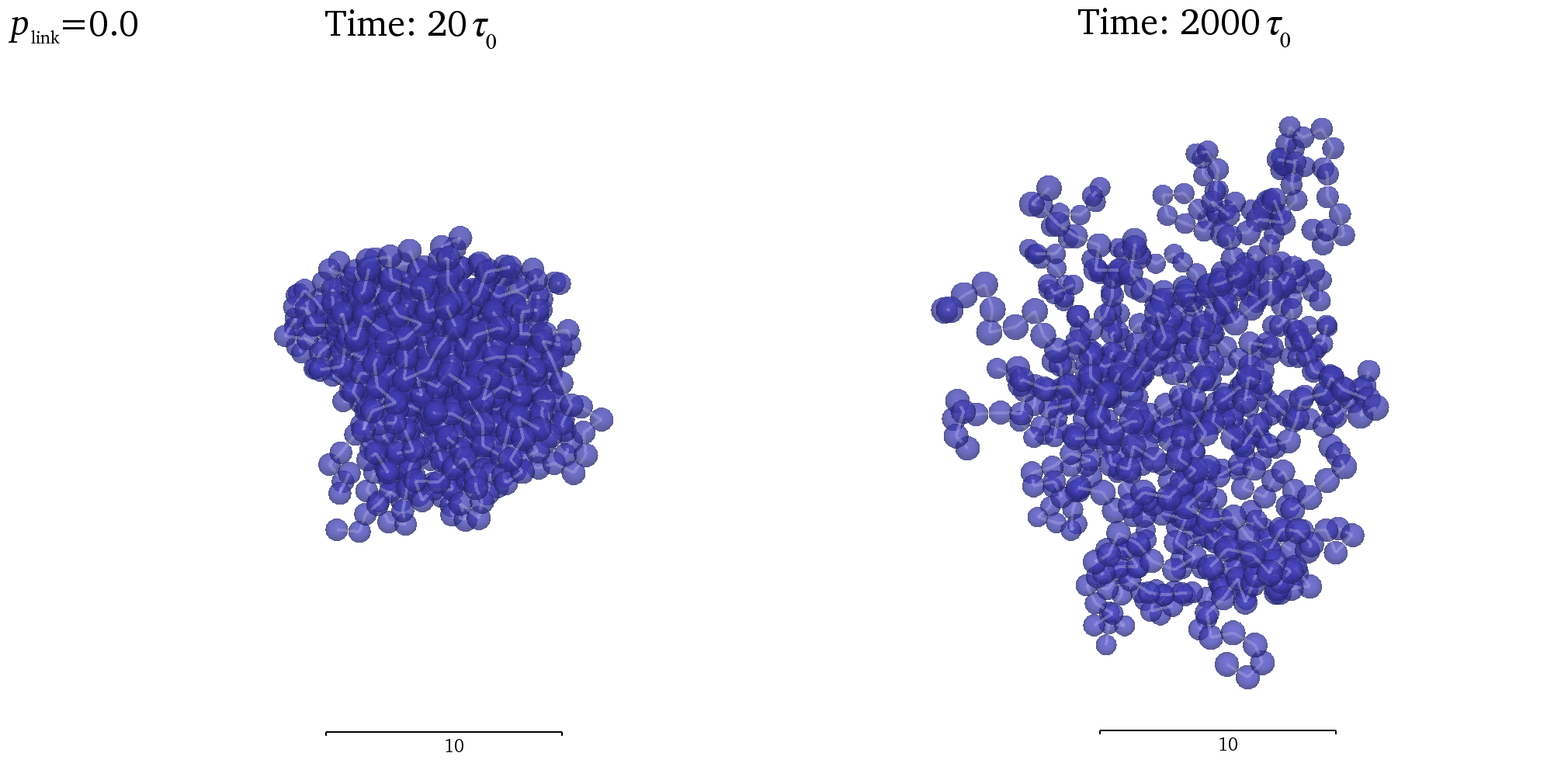}
				}
				\\
				\subfloat[]
				{
					\label{fgr:snapshots_p_link_0_4}
					\includegraphics[width=.95\textwidth]{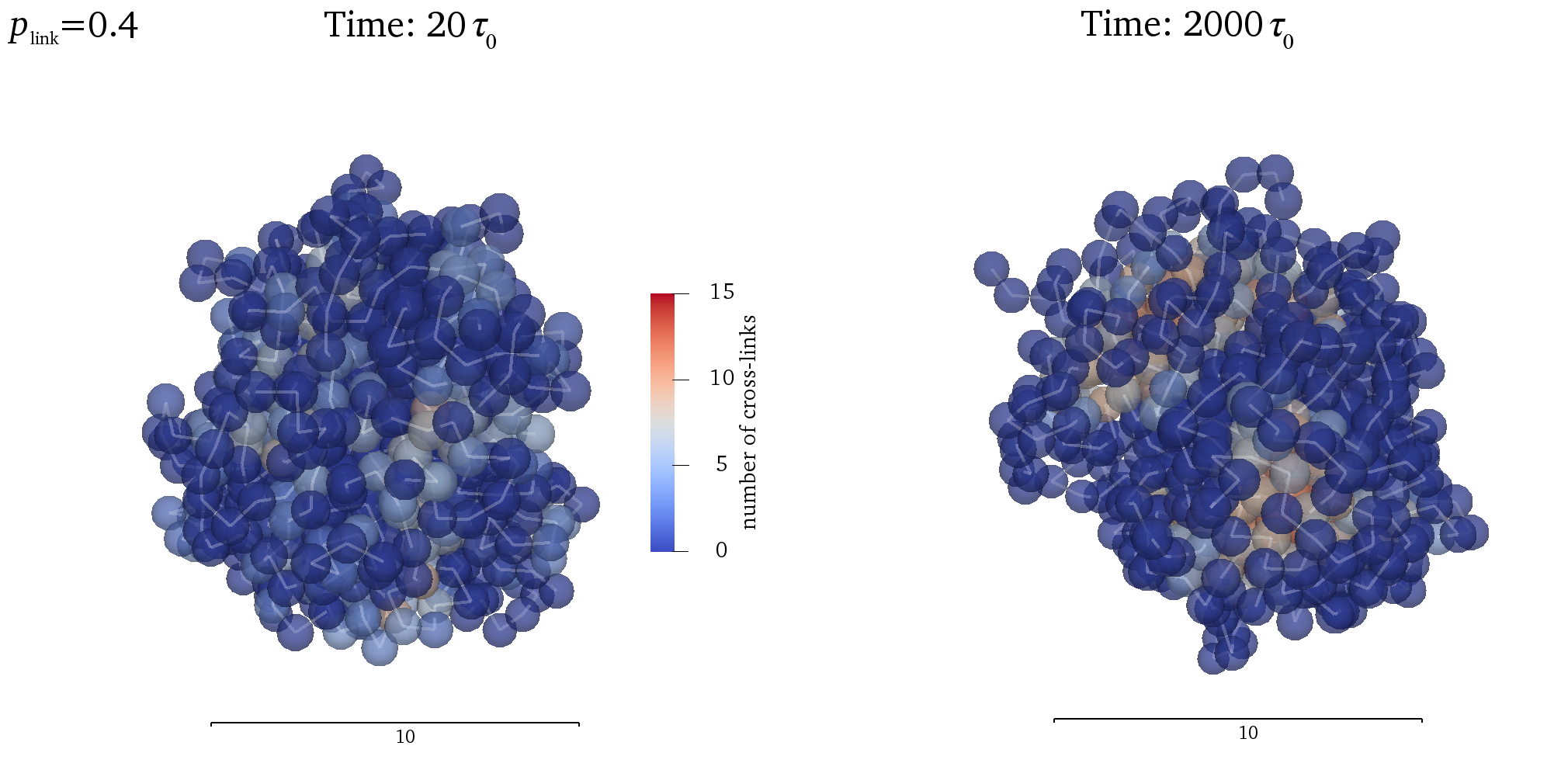}
				}
	
				\caption{Snapshots of the evolution of boluses in which no beads (\textbf{a}) or approximately 40\% of the beads (\textbf{b}) can form cross-links ($N_{\mathrm{bead}} = 20, \rho = 0.4, \epsilon = 5.0$). The beads are coloured according to the number of cross-links they participate in. The scale bar lengths are in units of $\sigma$. Note that for $p_{\mathrm{link}}=0$, the bolus size increases significantly with time as the molecules diffuse away. In contrast, the heavily cross-linked aggregate in \textbf{b} maintains a high density and an approximately constant size.}	
	
				\label{fgr:p_link_snapshots}
			\end{figure*}

			In Figure~\ref{fgr:p_link_snapshots}, we see snapshots of boluses at the beginning and the end of simulations. The polymeric molecules comprising the aggregates in Figure~\ref{fgr:p_link_snapshots}\subref{fgr:snapshots_p_link_0} can form no physical cross-links, whereas the fraction of linking beads in Figure~\ref{fgr:p_link_snapshots}\subref{fgr:snapshots_p_link_0_4} is $\approx 40\%$. As the comparison between these two cases indicates, introducing cross-linking interactions qualitatively changes the behaviour of polymeric aggregates, causing them to approach a stationary radius rather than disperse over time. The plots of the ensemble-averaged squared gyration radius $\langle R^2_{\mathrm{g \ bol}}(t) \rangle / \langle R^2_{\mathrm{g \ bol}}(0) \rangle$ in Figure~\ref{fgr:Rgbolsq_vs_t_diff_p_link_inset_dRgbolsqdt_vs_p_link} demonstrate that as $p_{\text{link}}$ is increased from 0, bolus behaviour transitions from dispersing over time to attaining a stationary gyration radius (see also Supplementary Movies 1 and 2). Looking at the bolus expansion rate $s$ shown in the inset of Figure 2 suggests that this is quite a sharp transition at  $p_{\text{link}}\approx 0.3$. This value is related to the fraction of linking beads required to bind all molecules in the aggregate together \textit{via} cross-links and is therefore related to a percolation threshold above which the entire aggregate is bound by cross-links. \par

			However, this is not a simple geometric percolation transition because it relates to the formation of a network of cross-links within a finite aggregate instead of an infinite medium. Moreover, the stability of the bolus is controlled not only by the number of cross-links in it but also by their strength, as well as additional factors, such as the initial volume fraction of polymer and the length of the polymeric chains. We discuss the most pertinent of these parameters below and choose the ensemble of boluses with  $p_{\mathrm{link}} = 8/N_{\text{bead}}, N_{\text{bead}} = 20, \rho = 0.4$  and $\epsilon = 5$ as a reference system since these parameters yield aggregates that are stable in quiescent conditions. \par

			\figdc{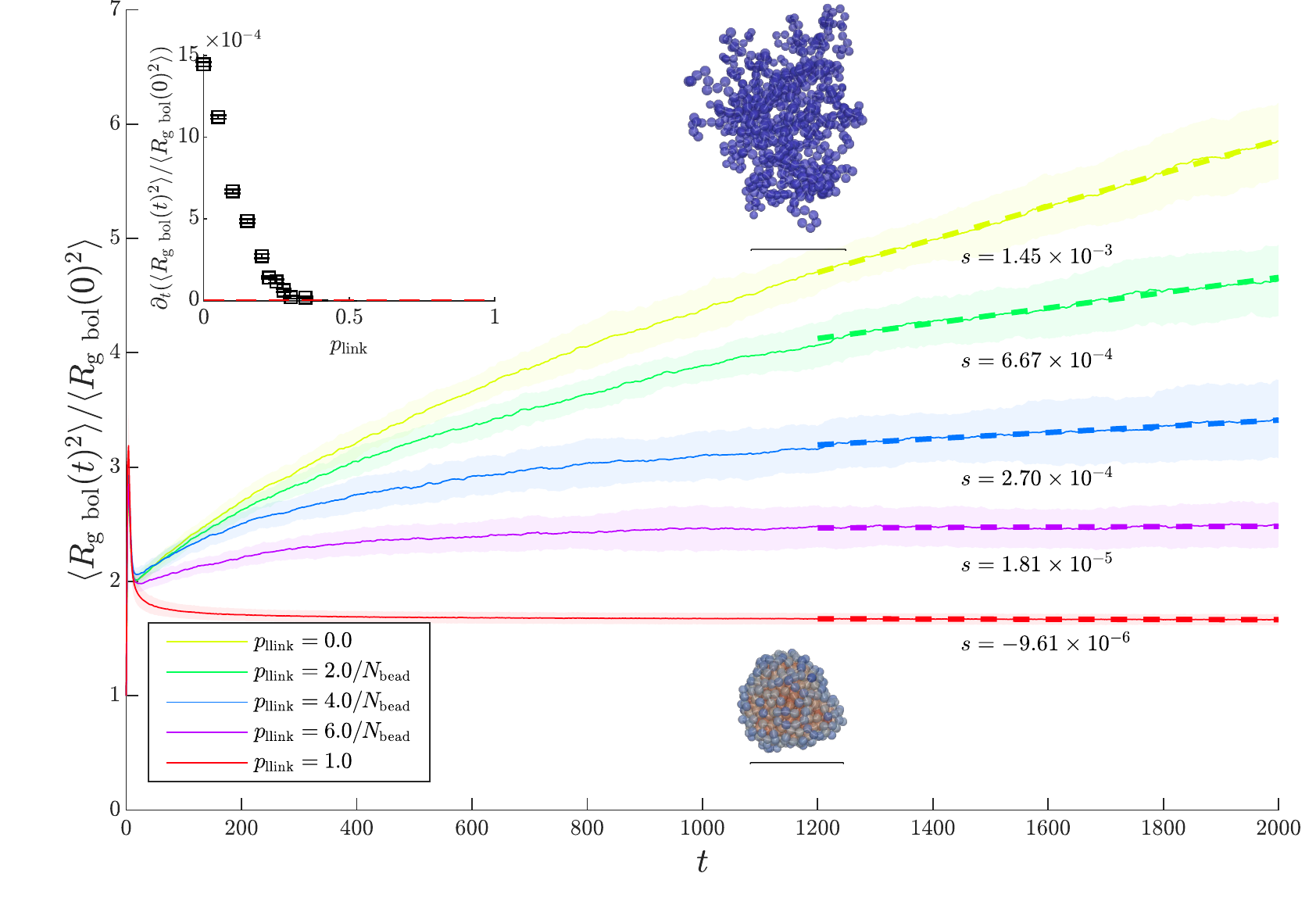}{{\bf The effect of linking probability.} Ensemble-averaged squared gyration radius $\langle R^2_{\mathrm{g \ bol}} \rangle$ vs. $t$ (in units of $\tau_0$), normalized by its value at $t = 0$ for boluses with fractions of linking beads ranging from 0 to 1 and $N_{\mathrm{bead}} =20, \rho =0.4, \epsilon =5.0$. The shaded areas indicate the sample standard deviation for each ensemble, and the dashed lines of slope $k$ are linear fits to the data for $R^2_{\mathrm{g \ bol \ k}}(t)/\langle R^2_{\mathrm{g \ bol}}(0) \rangle$ for individual boluses in each ensemble for the last 800 time units. Snapshots from simulations illustrate the two extreme cases ($p_{\mathrm{link}} = 0$ and 1) at $t = 2000 \tau_0$; note the scale bars of length $10\sigma$. Beads are coloured according to the number of cross-links they participate in with a scale that goes from blue to red as this number increases. 
			Inset - slopes of the fitted straight lines at various values of the fraction of linking beads in the bolus calculated from the data for $R^2_{\mathrm{g \ bol \ k}}(t)/\langle R^2_{\mathrm{g \ bol}}(0) \rangle$ for all boluses in each ensemble in the main plot and additional simulations at other $p_{\mathrm{link}}$. The dashed line corresponds to a zero slope, i.e., a zero surface expansion rate, $s = \partial_{t} (\langle R_{\mathrm{g \ bol}}^2(t)\rangle/\langle R_{\mathrm{g \ bol}}^2(t)\rangle) = 0$.}

		\subsubsection{Varying the interaction energy parameter $\epsilon$~~}
		
			In Figure~\ref{fgr:Rgbolsq_vs_t_diff_epsilon_inset_dRgbolsqdt_vs_epsilon} we show the long-time behaviour of the squared bolus gyration radius $\langle R^2_{\mathrm{g \ bol}}\rangle$ for different values of the interaction energy $\epsilon$. The figure demonstrates that merely having a linking bead fraction that ensures cross-links span the entire aggregate is not sufficient to make the bolus stable with respect to diffusion and that $\epsilon$ also needs to be above a threshold value for the aggregates to remain intact at long times. For our model system ($N_{\mathrm{bead}} = 20, \rho = 0.4, p_{\mathrm{link}} = 4.0/N_{\mathrm{bead}}$), we see that the long-time bolus expansion rate drops sharply at $\epsilon \approx  3$. 

			\figdc{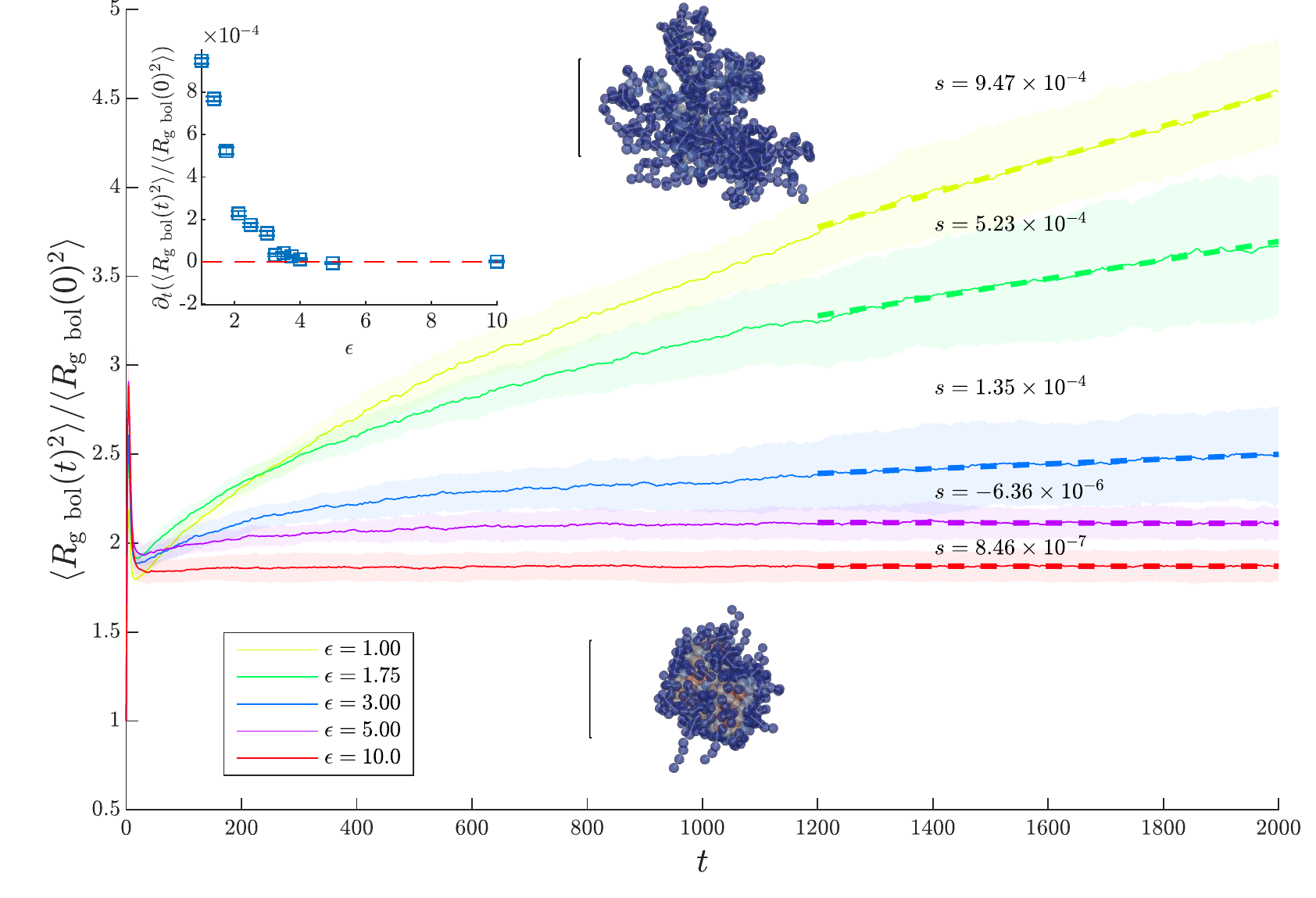}{{\bf The effect of the interaction energy.} Ensemble-averaged squared gyration radius $\langle R^2_{\mathrm{g \ bol}} \rangle$ vs. $t$ (in units of $\tau_0$), normalized by its value at $t = 0$ for boluses with different interaction energies $\epsilon$ and $N_{\mathrm{bead}} =20, \rho =0.4, p_{\mathrm{link}} =8.0/N_{\mathrm{bead}}$. The shaded areas indicate the sample standard deviation for each ensemble, and the dashed lines are linear fits to the data for $R^2_{\mathrm{g \ bol}}(t)/\langle R^2_{\mathrm{g \ bol}}(0) \rangle$ for all boluses in each ensemble for the last $800 \tau_0$. Snapshots from simulations illustrate the two extreme cases ($\epsilon = 1.00$ and 10.0) at $t = 2000 \tau_0$; the scale bars are of length $10\sigma$. Beads are coloured according to the number of cross-links they participate in and the colour map is the same for both snapshots. Inset - slopes of the normalized $R^2_{\mathrm{g \ bol}}(t)$ for all boluses in the individual ensembles long $t$ at various values of the intermolecular interaction energy parameter $\epsilon$. The slopes are calculated from fits to the data from the simulations in the main figure and ones at other $\epsilon$. The dashed line corresponds to a zero surface expansion rate, $s = \partial_{t} (\langle R_{\mathrm{g \ bol}}^2(t)\rangle/\langle R_{\mathrm{g \ bol}}^2(0)\rangle)$ = 0. }

		\subsubsection{Chemical breakdown~~}

			\figdc{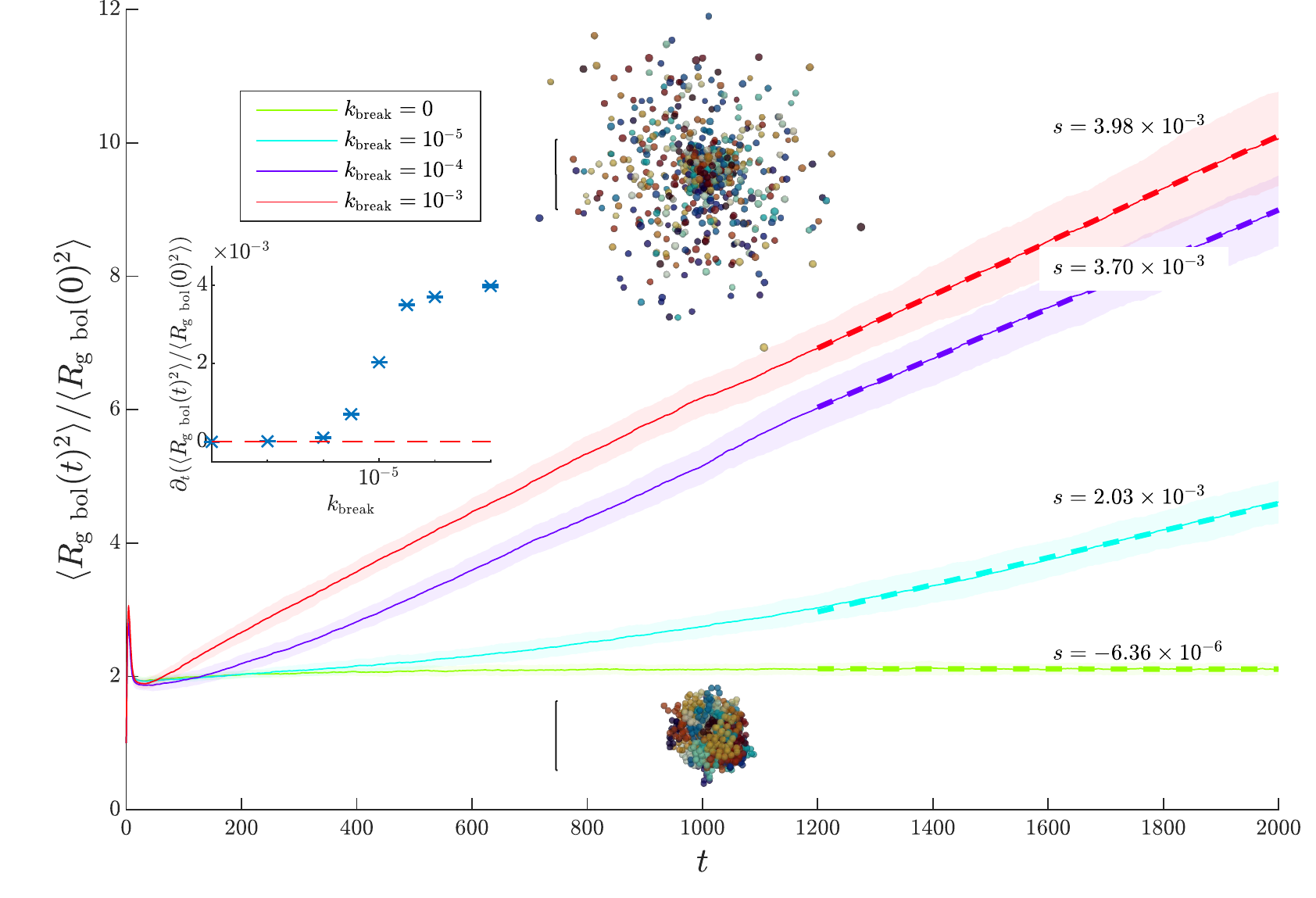}{{\bf The effect of chemical breakdown.} Ensemble-averaged squared gyration radius $\langle R^2_{\mathrm{g \ bol}} \rangle$ vs. $t$ (in units of $\tau_0$), normalized by its value at $t = 0$ for boluses with $N_{\mathrm{bead}} =20, \rho =0.4, p_{\mathrm{link}} =8.0/N_{\mathrm{bead}}, \epsilon =5.0$	 subjected to different rates of chemical breakdown $k_{\mathrm{break}}$. The shaded areas indicate the sample standard deviation for each ensemble, and the dashed lines are linear fits to the data for $R^2_{\mathrm{g \ bol}}(t)/\langle R^2_{\mathrm{g \ bol}}(0) \rangle$ for all boluses in each ensemble for the last $800 \tau_0$. The two snapshots taken at $t = t_{\text{sim}}$ illustrate the cases with the lowest and highest $k_{\mathrm{break}}$, with bead colours indicating the molecule to which beads belong at $t=0$; note the scale bars of length $10\sigma$.
			The inset shows the bolus surface expansion rates for long $t$ at various values of the chemical rate constant $k_{\mathrm{break}}$. The slopes are calculated from data for the boluses from the main figure and additional simulations at other $k_{\mathrm{break}}$.} \par

			We study the effect of polymer hydrolysis by varying the rate constant $k_{\mathrm{break}}$ over several orders of magnitude. We choose the values of $k_{\mathrm{break}}$ so that the number of bonds broken during the course of the simulations $t_{\text{sim}}$ varies between 0 and the total number of bonds in the bolus. 
			As seen in Figure~\ref{fgr:Rgbolsq_vs_t_for_diff_k_break_inset_dRgbolsqdt_vs_k_break}, which contains plots of the bolus ensemble-averaged squared gyration radius $\langle R^2_{\mathrm{g \ bol}}(t) \rangle$ at various $k_\mathrm{break}$, chemical breakdown in the model drives boluses that are stable in the absence of hydrolysis to disperse if $k_{\mathrm{break}} \gtrsim 10^{-6}$, corresponding to the cleavage of $\gtrsim 10\%$ of the bonds in the aggregate over the course of the simulation. \par

			The bolus evolution in the case of high $k_{\mathrm{break}}$ has an unexpected feature: in this case complete hydrolysis of the polymers occurs at $t \ll t_{\mathrm{sim}}$ and the bolus is reduced to a collection of individual beads. This allows the cross-link-forming beads to interact more strongly, forming a greater number of cross-links and thus a more tightly packed aggregate, than is possible when they participate in chemical bonds with non-linking beads. Thus, the breakdown of the aggregate is accompanied by the formation of a dense, heavily cross-linked core visible in the snapshots in Figure~\ref{fgr:Rgbolsq_vs_t_for_diff_k_break_inset_dRgbolsqdt_vs_k_break} and Supplementary Movie 3. One can envisage that such behaviour may occur if a co-polymer containing hydrophobic and hydrophilic monomers is broken down at a high rate into small clusters in water, leading hydrophilic clusters to dissolve in the solvent and hydrophobic ones to aggregate. \par

	\subsection{Boluses in flow}

		Here, we induce a simple shear flow by introducing moving horizontal solid walls to the system as described in Section \ref{Methods} and investigate how it affects bolus dynamics both with and without simultaneous chemical breakdown of the polymer molecules.

		\subsubsection{Tumbling, tank-treading and breakdown in shear flow~~}

			Boluses under shear exhibit three different regimes depending on the imposed shear rate $\dot{\gamma}$. For boluses that are stable under quiescent conditions, we nondimensionalize $\dot{\gamma}$ with a characteristic time $\langle t_{\text{bolus}}\rangle$, which we define as the time required for $\langle R^2_{\mathrm{g \ bol}}(t) \rangle$ of a bolus with identical parameters to reach 95\% of $\langle R^2_{\mathrm{g \ bol}}(t_\mathrm{sim}) \rangle$ in the absence of shear. Note that, as before, we disregard the initial period in which $\langle R^2_{\mathrm{g \ bol}}(t) \rangle$ goes through a maximum. For this reason, when determining $\langle t_{\text{bolus}}\rangle$, we start measuring $\langle R^2_{\mathrm{g \ bol}}(t) \rangle$ at $t_0 = t_{\mathrm{sim}}/20$. The dimensionless bolus Weissenberg number is then $Wi = \dot{\gamma}\langle t_{\text{bolus}}\rangle$, with $\langle t_{\text{bolus}}\rangle > t_0$; for our model system ($N_{\mathrm{bead}} = 20, \rho = 0.4, p_{\mathrm{link}} = 4.0/N_{\mathrm{bead}}, u_{\mathrm{wall}} = 0, k_{\mathrm{break}} = 0$), $\langle t_{\text{bolus}}\rangle = 140 \tau_0$.  \par		

			\begin{figure*}
				\centering
				\subfloat[]
				{
					\label{fgr:snapshots_Wi_0_6}
					\includegraphics[width=.6\textwidth]{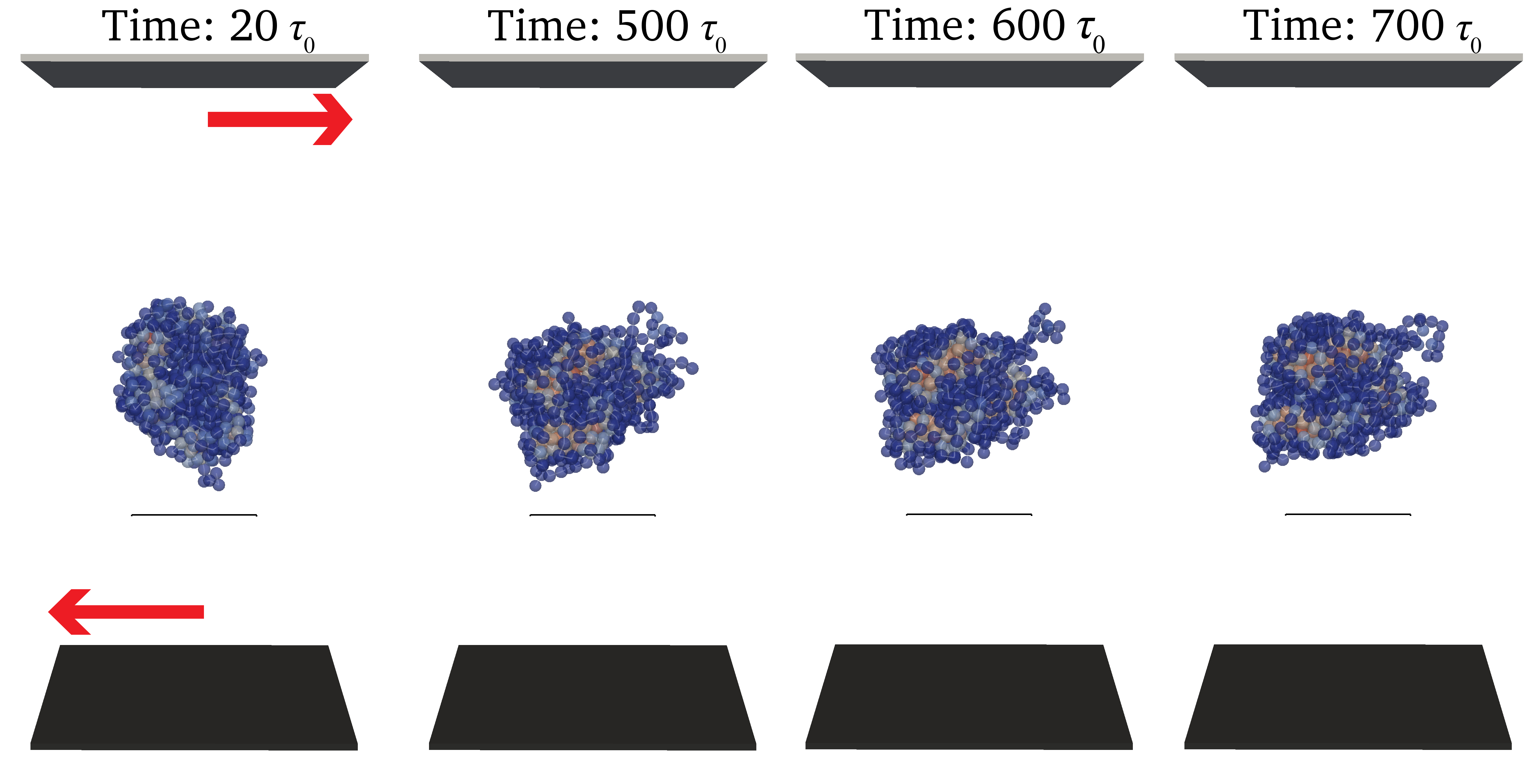}
				}
				\\
				\subfloat[]
				{
					\label{fgr:snapshots_Wi_5_8_tumbling}
					\includegraphics[width=.6\textwidth]{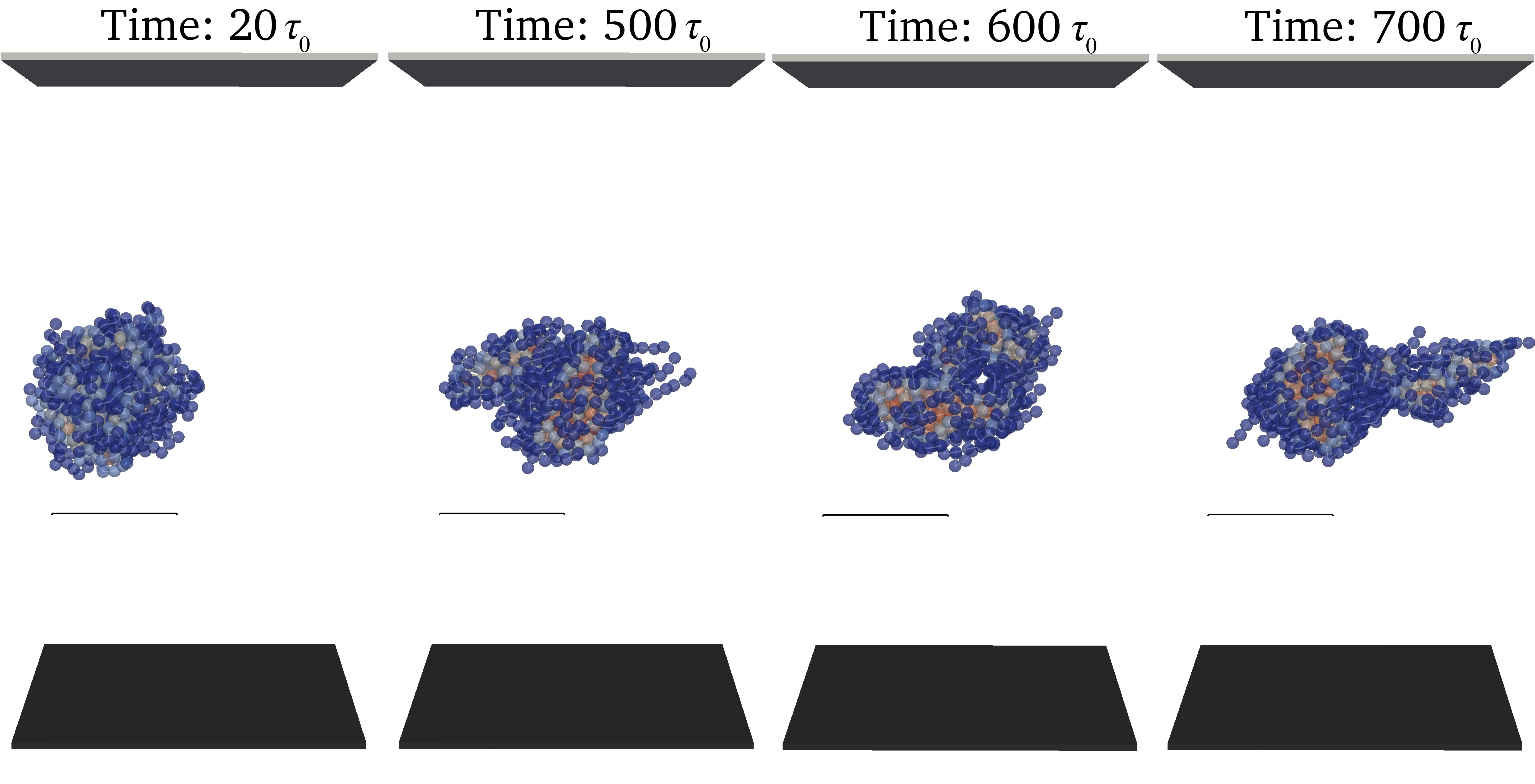}
				}
				\\
				\subfloat[]
				{
					\label{fgr:snapshots_Wi_29}
					\includegraphics[width=.8\textwidth]{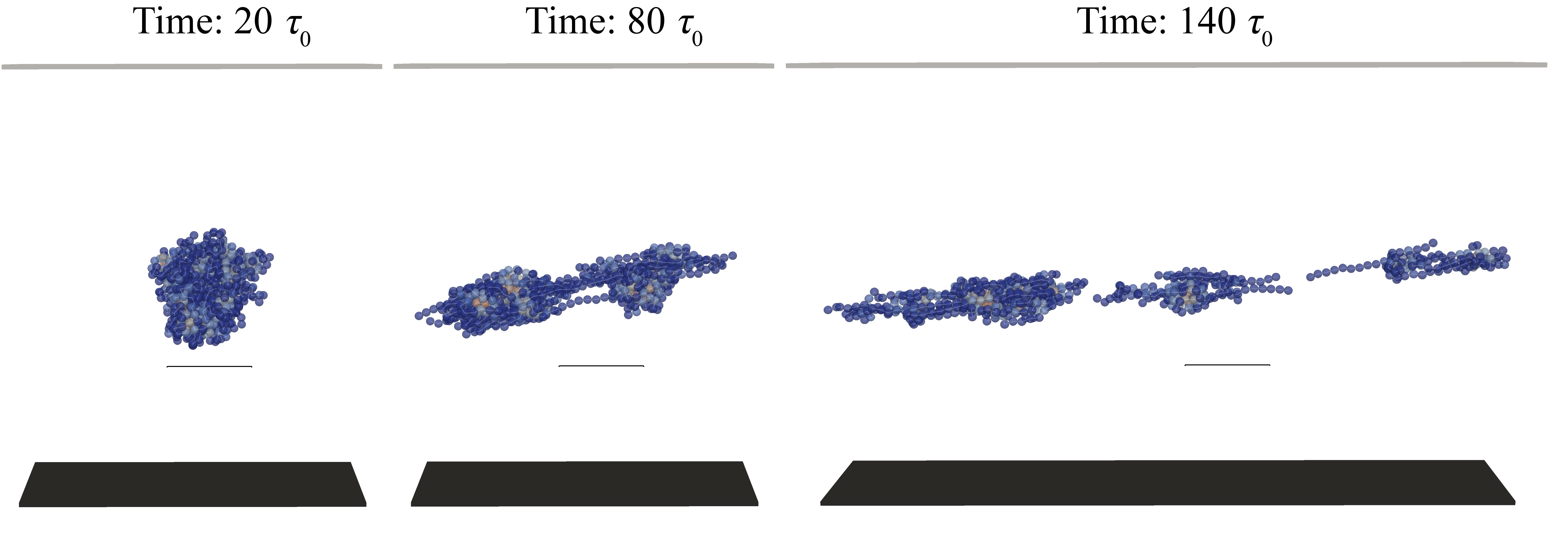}
				}
	
				\caption{Snapshots of the evolution of boluses with $N_{\mathrm{bead}} = 20, \rho = 0.4, p_{\mathrm{link}} = 8/N_{\mathrm{bead}} = 20, \epsilon = 5.0$ subjected to flows with different shear rates. The shear is created by the motion of the horizontal solid walls as indicated by the arrows in the first frame in \textbf{a}. The scale bars are of length $10\sigma$. \textbf{a}. Regime i) - at low $Wi$ ($Wi = 0.58$), the bolus is not significantly affected by the flow. \textbf{b}. Regime ii) - at intermediate $Wi$ ($Wi = 5.8$), the bolus performs a tumbling/tank-treading motion. \textbf{c}. Regime iii) - at high $Wi$ ($Wi = 29$), the bolus is broken apart by the flow.}
	
				\label{fgr:Wi_snapshots}
			\end{figure*}

			We now discuss the different regimes of bolus behaviour in shear flow. i) At $Wi \ll 1$, the bolus is not significantly perturbed by the shear flow and both $\langle R^2_{\mathrm{g \ bol}}(t) \rangle$ and $s$ are unchanged with respect to the quiescent case (Figure~\ref{fgr:Wi_snapshots}\subref{fgr:snapshots_Wi_0_6}). ii) At intermediate $Wi$, $Wi \sim 1$, boluses are deformed and move collectively in the direction of the flow (Figure~\ref{fgr:Wi_snapshots}\subref{fgr:snapshots_Wi_5_8_tumbling}, Supplementary Movie 4). Individual polymer chains break off from some of the boluses in an ensemble (see Supplementary Movie 5), but overall, the bolus maintains its structure. The transition between regimes i) and ii) occurs when the energy dissipated due to viscous friction over the characteristic time for the bolus $\langle t_{\text{bolus}}\rangle$ becomes much larger than the total energy of the cross-links in it. iii) At $Wi \gg 1$, the boluses are broken apart by the shear flow (Figure~\ref{fgr:Wi_snapshots}\subref{fgr:snapshots_Wi_29}, Supplementary Movie 6). \par

			The final bolus size $\langle R_{\mathrm{g \ bol}}(t_{\mathrm{sim}})\rangle/\langle R_{\mathrm{g \ bol}}(0)\rangle$ for $Wi\gg 1$ is orders of magnitude greater than the box length $L$. This means that the box no longer adequately simulates an infinite medium and that intermolecular interactions are stronger than they would be in an infinite box, and that the results we report for this case at long times are approximate; the same is true of simulations in which individual molecules are separated from the bolus by the shear flow. However, we expect that the finite size of the box does not introduce a significant error in the quantities of interest to us in the latter case because the splitting of individual molecules from an aggregate of $\sim 30$ chains has a minor effect on its gyration radius.

			In regime ii), in which the aggregates move approximately as solid bodies, there are two limiting modes of motion that the bolus can exhibit: tumbling and tank-treading. In the first of these, tumbling, the polymeric chains experience large conformational changes and alternate between stretched and collapsed states. In the second one, tank-treading, individual beads rotate about the bolus centre of mass and the conformation of the chains is approximately constant. 			
			See Refs.~\cite{Chen2013,Formanek2019} for discussion of these modes for individual ring polymers and nanoparticles consisting of a single polymeric chain, respectively.\par

			We use two correlation functions to describe this collective motion, on the scale of the entire bolus, at intermediate shear rates. The first one is the cross-correlation function of the diagonal components of $G$ in the flow and gradient direction, which characterizes tumbling,
			\begin{equation}\label{eq14}
				C_{xz} = \frac{\langle \updelta G_{zz\text{ k}}(t_0)\updelta G_{xx\text{ k}}(t) 	\rangle}{\sigma_{G_{zz}(t_0)}\sigma_{G_{xx}(t_0)}},
			\end{equation} 
			where 
			\begin{equation}\label{eq15}
				\updelta G_{\text{ii k}}(t) = G_{\text{ii k}}(t) - \langle G_{\text{ii}}(t) \rangle
			\end{equation}
			and
			\begin{equation}\label{eq16}
				\sigma_{G_{\text{ii}}(t)} =\sqrt{\frac{N_{\text{ens}}}{N_{\text{ens}}-1} \left( \langle G^2_{\text{ii} }(t) \rangle - \langle G_{\text{ii}}(t) \rangle^2 \right) }
			\end{equation}
			are the standard deviations of the diagonal components of the gyration tensor. We choose the offset time $t_0 = t_{\mathrm{sim}}/20$ so that the repulsion-dominated initial period during which the bolus gyration radius goes through a maximum does not contribute to the correlation functions. \par
			Negative peaks in the cross-correlation function $C_{xz}$ are a hallmark of tumbling motion~\cite{Formanek2019}. These peaks arise because the polymer chains are preferentially stretched along the flow direction, but thermal fluctuations cause stretching in the gradient direction. This causes the chains to contract and subsequently extend along the flow.\par
			The correlation function that characterizes tank-treading is
			\begin{equation}\label{eq17}
				\begin{split}
					C_{\text{angle}}(t) = \frac{\langle \sin\left(2\beta(t_0)\right)\sin\left(2\beta(t)\right) \rangle}{\langle \sin\left(2\beta(t_0)\right)^2 \rangle} = \\ \frac{\left< \sum_{\text{i}=1}^{N_{\text{poly}}}{\sum_{\text{j}=1}^{N_{\text{bead}}}{ \sin\left(2\beta_{\text{ijk}}(t_0)\right)\sin\left(2\beta_{\text{ijk}}(t)\right)}} \right>}{\left\langle \sum_{\text{i}=1}^{N_{\text{poly}}}{\sum_{\text{j}=1}^{N_{\text{bead}}}{\sin\left(2\beta_{\text{ijk}}(t_0)\right)^2 }} \right\rangle},
					\end{split}
			\end{equation}
			where $\beta_{\mathrm{ijk}}$ is the angle between the vector connecting the bead with position vector $\bm{r}_{\text{ijk}}$ to the bolus centre and the instantaneous first principal component of the bolus bead positions. The principal components of the instantaneous set of bead positions are basis vectors defined through a linear transformation of the Cartesian basis in which we record the positions. The transformation is defined such that the principal components maximize variance and are orthogonal to each other \cite{Jolliffe2002}. The principal components correspond to the axes of an ellipsoid fit to the distribution of bead positions, and the first principal component corresponds to this ellipsoid's principal axis along which statistical variation is greatest \cite{Jolliffe2002}, thus defining the main bolus axis.
			$\beta_{\mathrm{ijk}}$ is therefore  defined through
			\begin{equation}\label{eq18}
				\cos(\beta_{\text{ijk}}(t)) = \frac{\left(\bm{r}_{\text{ijk}}-\bm{r}_{\text{av k}}(t) \right)\cdot\textbf{\~{e}}_{\text{p}}}{\abs{\bm{r}_{\text{ijk}}-\bm{r}_{\text{av k}}(t)}},
			\end{equation}
			where $\textbf{\~{e}}_{\text{p}}$ is a unit vector in the direction of the first principal component.
			
			Damped oscillations of $C_{\text{angle}}$ with time are a characteristic sign of tank-treading motion \cite{Chen2013,Formanek2019}. In contrast with tumbling, which causes polymer molecules to alternate between stretched and compressed conformations, tank-treading motion occurs with rotation of the individual beads around the bolus centre of mass that maintains the conformation of individual molecules approximately constant~\cite{Formanek2019}.  \par

			\begin{figure*}
				\centering
				\subfloat[]
				{
					\label{fgr:Cxz_diff_Wi}
					\includegraphics[width=8cm]{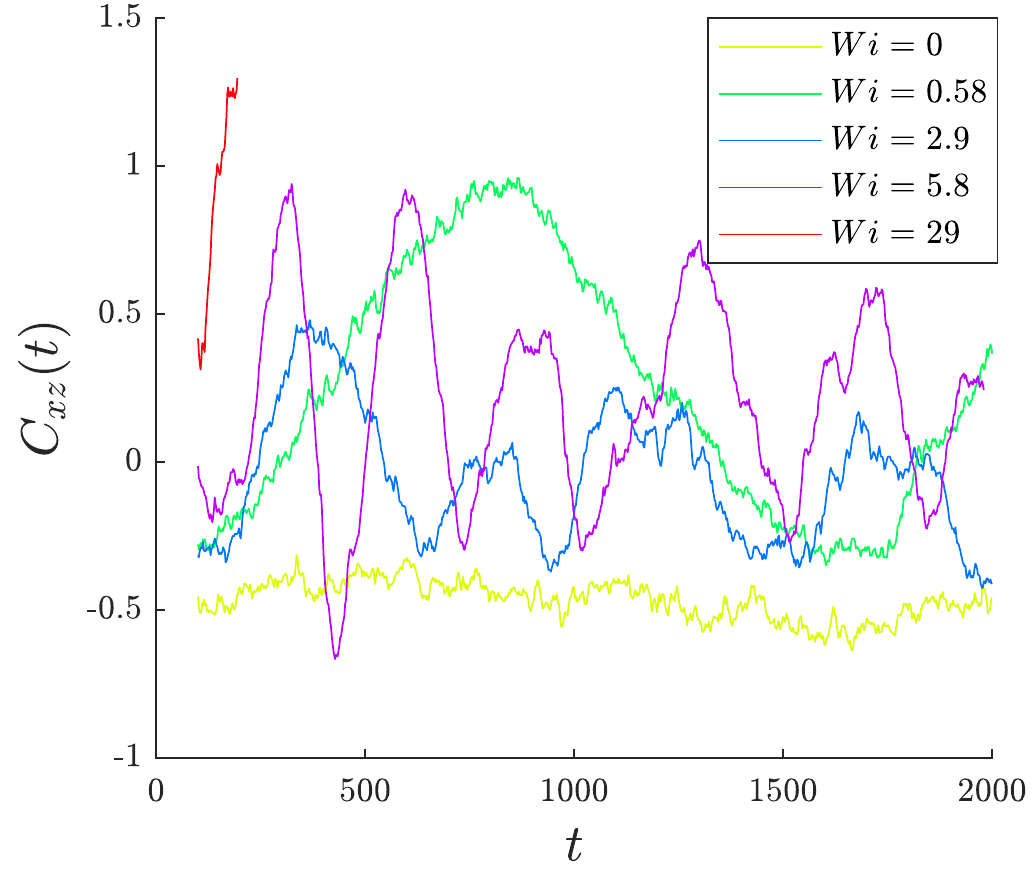}
				}
				\\
				\subfloat[]
				{
					\label{fgr:C_angle_diff_Wi}
					\includegraphics[width=8cm]{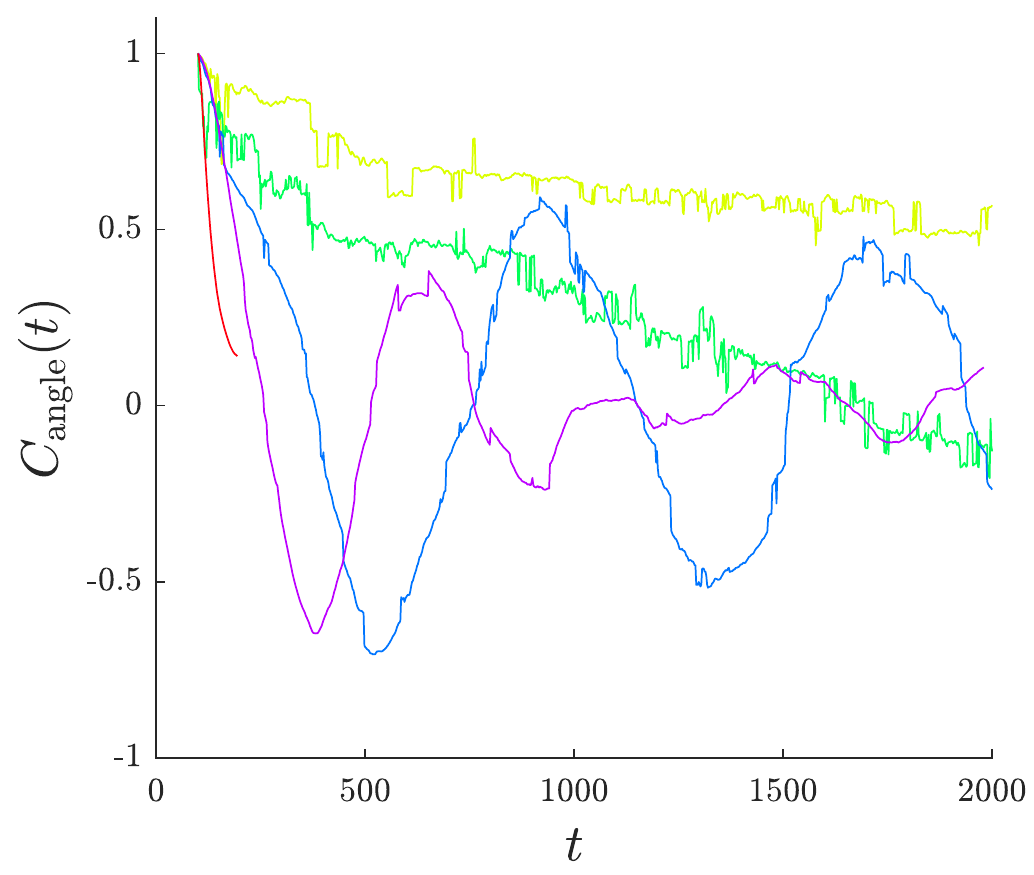}
				}	
				\caption{
					Correlation functions characterizing tumbling and tank-treading versus $t$ in units of $\tau_0$ plotted up to the moment when $\langle R^2_{\mathrm{g \ bol}}(t_{\mathrm{sim}})\rangle = (L/2)^2$; at $Wi = 29.2$ this happens at $t \approx 200 \tau_0$ (red curves).
					\textbf{a}. Correlation function $C_{xz}(t)$ for boluses with $N_{\mathrm{bead}} =20, \rho =0.4, p_{\mathrm{link}} =8.0/N_{\mathrm{bead}}$, and $\epsilon =5.0$ subjected to shear flows with different $Wi$. The negative peaks in $C_{xz}(t)$ at $Wi\sim 0.58-5.8$ indicate anti-correlation between the deviations from the average for the bolus lengths in the flow ($\updelta G_{zz\ \mathrm{k}}(t)$) and gradient ($\updelta G_{xx \ \text{k}}(t)$) directions. This demonstrates that the motion of boluses at $Wi \approx 3$ (blue) and $Wi \approx 6 $ (purple) includes contributions from tumbling. The boluses at $Wi \approx 0.6 $ (green) also exhibit two shallow negative peaks indicative of slow tumbling. Note the qualitative difference with the correlation functions for boluses in a quiescent fluid (yellow).					\textbf{b}.Correlation function $C_{\text{angle}}(t)$ for boluses subjected to shear flows with different $Wi$. Note the signature damped oscillations of tank-treading for $Wi = 2.9$ and $5.8$ (blue and purple, respectively), and the qualitatively different behaviour of $C_{\text{angle}}(t)$ for other values of $Wi$.
				}
				\label{fgr:Corr_functions}
			\end{figure*}

			These correlation functions are plotted in Figures~\ref{fgr:Corr_functions}\subref{fgr:Cxz_diff_Wi} and~\ref{fgr:Corr_functions}\subref{fgr:C_angle_diff_Wi}. They indicate that, as $Wi$ is increased from 0, the boluses studied here first exhibit slow tumbling motion at $Wi~\sim~1$ (green curve, Figure~\ref{fgr:Corr_functions}\subref{fgr:Cxz_diff_Wi}), and then move in a way that combines tumbling and tank-treading (blue and purple curves, Figure~\ref{fgr:Corr_functions}) at higher $Wi$.  
			
		\subsubsection{Synergy of shear flow and chemical breakdown~~}
		
\begin{figure}
				\centering
				\includegraphics[width=8.5cm]{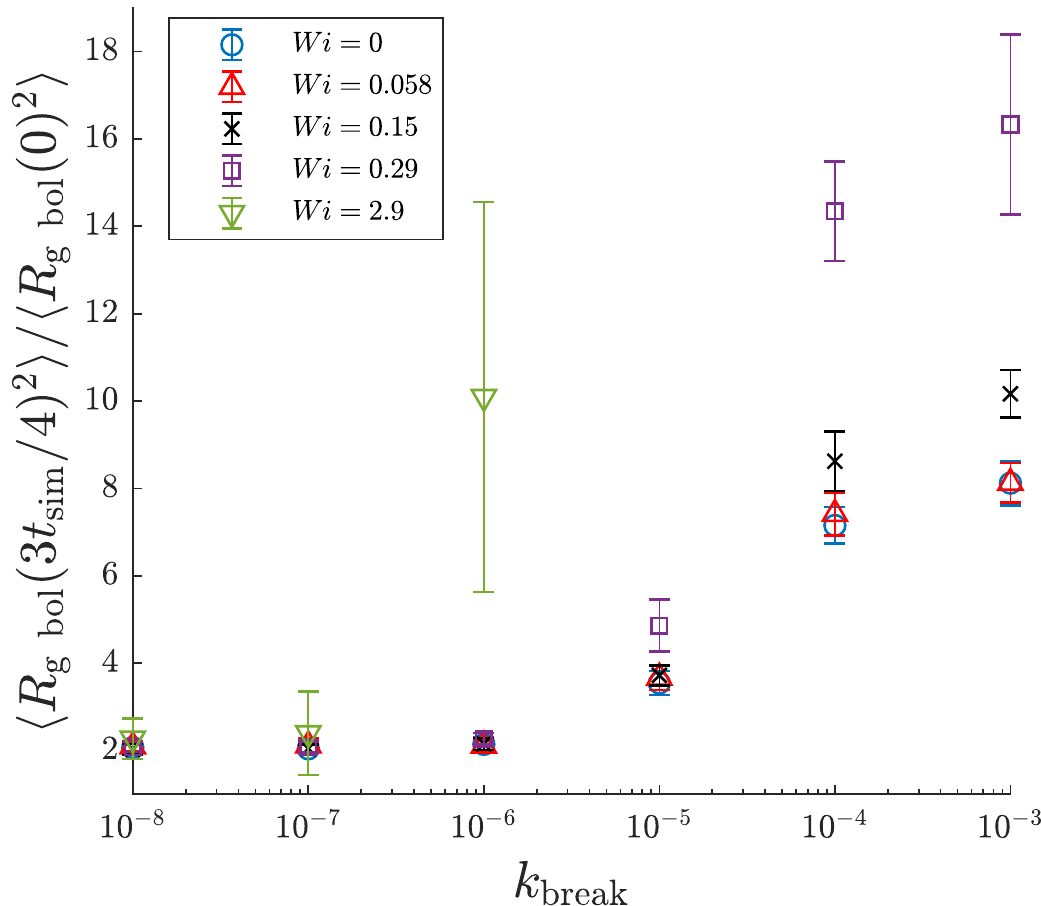}
				\caption{Ensemble-averaged squared gyration radius at $t=3t_{\text{sim}}/4 = 1500 \tau_0$, normalized by its value at $t = 0$ versus $k_{\mathrm{break}}$ in the absence of shear flow (circles), at $Wi = 0.06$ (triangles), $0.15$ (crosses), $0.29$ (squares), and $2.9$ (inverted triangles). Notably, Weissenberg numbers lower than unity significantly enhance bolus dispersion at high chemical reaction rates although even a tenfold increase of $Wi$ has little effect on aggregate size in the absence of hydrolysis or sufficiently low reaction rates (compare the data for $Wi = 0, 0.15 $ and $2.9$ at $k_{\mathrm{break}} = 10^{-8}$). The squared gyration radii at $k_{\mathrm{break}} \geq 10^{-5}$ are not shown for the last set of points because they exceed the others by orders of magnitude, ranging between $\sim 100$ and $\sim 900$. At long times and comparatively high $Wi$ and $k_{\mathrm{break}}$, a significant fraction of the beads reach distances greater than $L/2$ from the centre of the simulation box, which means that the latter is not large enough to effectively simulate an infinite medium for these beads. Therefore, we plot $\langle R^2_{\mathrm{g \ bol}}(t) \rangle$ at $t=3t_{\text{sim}}/4$, which is approximately when the finite-size effects start to play a role for $Wi = 0.15$ and $k_{\mathrm{break}} = 10^{-3}$; this value of $Wi$ approximately corresponds to the upper limit for the shear rate in the stomach (see Discussion). The sizes we report for $k_{\mathrm{break}} = 10^{-6}$ at $Wi = 2.9$ and $k_{\mathrm{break}} \geq 10^{-4}$ at $Wi = 0.29$ are approximate as they do not account for the finite size of the box.}
				\label{fgr:Rgbolsqfinal_vs_k_break_shear}
			\end{figure}

			Finally, we consider the case of bolus evolution in the presence of both shear flow and chemical breakdown, which is particularly relevant to modelling the digestive tract where muscle contractions induce mixing and enzymes catalyse hydrolytic reactions. Figure~\ref{fgr:Rgbolsqfinal_vs_k_break_shear}, which contains data on $\langle R_{\mathrm{g \ bol}}(k_{\mathrm{break}})\rangle$ at different dimensionless shear rates, illustrates the synergistic effect of the two factors. The graph demonstrates that the combination of fast polymer hydrolysis and a low Weissenberg number is more efficient than either of the two on its own (see also Supplementary Movie 7).  \par

			Figure~\ref{fgr:Rgbolsqfinal_vs_k_break_shear} shows that flows with $Wi = 0.29$,   which in the absence of chemical breakdown have a negligible effect on bolus size lead to a considerable increase in the squared gyration radius at long times and $k_{\mathrm{break}}$ high enough to cleave a substantial fraction of the chemical bonds in the aggregate ($\langle R_{\mathrm{g \ bol}}^2 \rangle$ increases by $\approx 30\%$ at $k_{\mathrm{break}} = 10^{-5}$ approximately twofold at $k_{\mathrm{break}} = 10^{-4}-10^{-3}$). \par

			The reason such low shear rates affect boluses only if the latter undergo a hydrolytic reaction is that fast hydrolysis generates a substantial fraction of individual beads which are then much more easily dispersed by the flow than long polymeric chains would be. Similar effects may play a role in digestion. \par
				
\section{Discussion}

	Our simulations of the dynamics of physically cross-linked aggregates consisting of linear polymers (boluses) provide insight into the process of digestion at the mesoscale. The coarse-grained models we employ allow us to pinpoint the key factors controlling the breakdown of such aggregates. We demonstrate that in a quiescent fluid, the stability of polymeric boluses with respect to diffusion is mainly controlled by the fraction of individual beads within them that can form physical cross-links ($p_{\mathrm{link}}$) and the energy of the cross-linking interactions ($\epsilon$). The bolus surface expansion rate $s$ sharply decreases to zero at a value of $p_{\mathrm{link}}$ that corresponds to a network of cross-links encompassing the entire aggregate provided that the cross-links are strong enough to hold the molecules together, i.e., that $\epsilon$ is above a threshold value. \par

	Two major factors control the breakdown of biopolymers in the digestive tract - enzymatic hydrolysis and shear flow. Our simple model of enzymatic polymer hydrolysis posits that all chemical bonds within the bolus degrade at random with the same probability. It demonstrates that a rate of hydrolysis which cleaves a significant fraction of the bonds within an aggregate over the simulated time period causes boluses that are stable in the absence of hydrolysis to disperse. \par

	We also investigate the effect of simple shear on polymeric aggregates by introducing two horizontal solid walls moving in opposite directions. We observe that aggregates exhibit three different regimes depending on the magnitude of the Weissenberg number $Wi$ for the flow:  i) At $Wi \ll 1$, the flow does not have an appreciable effect on the aggregates. ii) At $Wi \sim 1$, boluses move collectively along the flow in a regime that combines tumbling and tank-treading, and individual molecules may detach from the main aggregates. iii) At $Wi \gg 1$, the shear flow is sufficiently strong to break the aggregates down completely. \par

	Finally, we show that when combined, shear flow and chemical breakdown of the polymers act in synergy to disperse the aggregates. In the limit of fast chemical breakdown in which the polymeric chains are completely split into individual beads over the duration of the simulation, flows with $Wi \sim 0.1$ significantly aid bolus dispersion. This is in contrast to the case of aggregates in the absence of hydrolysis in which a shear rate of the same order of magnitude would at most induce slow tumbling. Such low $Wi$ have an appreciable effect at high hydrolysis rates because the flow need only disperse individual beads rather than long polymeric chains. \par

	Let us now discuss how our mesoscale model system might be approximately mapped to physical units. We first assume that a single simulated chain should be mapped to a single biopolymeric molecule, i.e. we set the contour length $L_{\mathrm{c}} = N_{\text{bead}}\sigma$ equal to that of biopolymers that occur in the human diet. First, we take the example of the linear carbohydrate amylose, which is a major component of starch, and thus essential to human nutrition. We set $T$ to $\SI[mode=text]{25}{\celsius}$ and use the experimental contour length $L_{\mathrm{c}} = \SI{270}{nm}$ and linear density of $\rho^\mathrm{L}= \SI{2.39e-12}{kg.m^{-1}}$ of amylose~\cite{Stokke1987}. For a polymer consisting of 20 coarse-grained beads, this means that $\sigma = \SI{13.5}{nm}$ and $R_{\text{sphere}} = \SI{81}{nm}$. If instead we take the example of the much longer biopolymer xanthan ($L_{\mathrm{c}} = \SI{2.608}{\micro\meter}$ in its double-stranded form,   $\rho^\mathrm{L}= \SI{3.32e-12}{kg.m^{-1}}$~\cite{Stokke1987}) which is a common gelling agent in the food industry, we get $\sigma = \SI{0.13}{\micro\meter}$ and $R_{\text{sphere}}= \SI{0.78}{\micro\meter}$.
		
	For comparison, image analysis of rice-based gastric digesta from pigs shows a particle distribution with areas ranging from $10^{-9}$ to $\SI{3.5e-5}{\meter^2}$ \cite{Bornhorst2013a}, corresponding to radii between $\sim \SI{30}{\micro\meter}$ and $\sim \SI{6}{\milli\meter}$, i.e., if we map our model polymers to xanthan, the size of the aggregates we study is about an order of magnitude smaller than the smallest particles in digesta. \par

	In this mapping the range of $Wi$ we study corresponds to shear rates of $\sim 10^{4}$ to $\sim 10^{7}$ $\mathrm{s^{-1}}$ for amylose and $\sim 10^{3}$ to $\sim 10^{6}$ $\mathrm{s^{-1}}$ for xanthan. Shear rates in the digestive tract have not been accurately measured~\cite{Dikeman2006} but based on simulation data~\cite{Ferrua2011}, we can estimate them to be $\sim\SI{1}{\s^{-1}}$. Our estimates predict that the shear rates required to perturb the aggregates formed from amylose and xanthan are much higher than this value, implying that mixing in the digestive tract would play no role in the breaking down of boluses with such small dimensions ($\sim\SI{0.1}{\micro\meter}$ and $\sim\SI{1}{\micro\meter}$, respectively).  \par

	An alternative way of interpreting the model is to equate the initial aggregate size, $R_{\text{sphere}} = 6\sigma$, to the approximate radius of the smallest particles in digesta observed by Bornhorst et al.~\cite{Bornhorst2013a}, $R_{\text{sphere}} \approx \SI{32}{\micro\meter}$. This is equivalent to assuming that each polymer in the bolus represents multiple entangled polymeric chains. Taking the same linear density as amylose, $\rho^\mathrm{L}= \SI{2.39e-12}{kg.m^{-1}}$~\cite{Stokke1987}, we find that the range of shear rates we explore falls between $5$ and $\SI{5e3}{s^{-1}}$, and that the bead diameter is $\sigma = \SI{5.3}{\micro\meter}$. In this mapping, given the presence of hydrolysis, a physiologically relevant shear rate of $\dot{\gamma} \approx \SI{11}{\s^{-1}}$ ($Wi = 0.15$) is sufficient to significantly affect the digestion of boluses. \par

	The coarse-grained mesoscale model discussed in this paper provides insight into generic mechanisms of polymer aggregate breakdown relevant to modelling the process of digestion. Our work considers some of the main factors at play in the digestive tract, but there are many simplifying assumptions that require further investigation. The physical cross-links present in the model can serve as an approximation of the hydrogen bonds that act between starch molecules \cite{Liang2015}. 
	However native starch consists of branched-chain amylopectin and linear amylose molecules \cite{Buleon1998} which can be	organized in complex structures known as granules which vary between 1 and $\sim \SI{100}{\micro\meter}$ in size \cite{Buleon1998} and contain alternating amorphous and crystalline layers, although these are typically disrupted during food processing \cite{Zhang2014a} and further perturbed during digestion. First steps towards modelling more realistic polymer architectures could be to consider chain branching, polydispersity or the effects of electrostatic interactions. \par

	In our model for polymer hydrolysis the chemical bonds break down spontaneously at a constant rate regardless of their position within the bolus. An enzyme that hydrolyses amylose according to this mechanism is $\upbeta$-amylase, which is found in plants, see e.g.~Bird and Hopkins \cite{Hopkins1954}. A more realistic model of hydrolysis due to the $\upalpha$-amylase present in human saliva might follow a multiple attack mechanism in which the enzyme at first binds to a random site along the carbohydrate chain and then hydrolyses several bonds before detaching from it \cite{Robyt1967}. Moreover, being macromolecules themselves, enzymes diffuse through the polymeric aggregates at a finite rate and a more detailed model should account for enzyme diffusion. In this case, chains at the surface of the aggregate would be attacked first; individual linking beads would have more time to diffuse after hydrolysis and would be less likely to form a dense core as they do in our simulations (see the snapshot for high $k_{\mathrm{break}}$ in Figure~\ref{fgr:Rgbolsq_vs_t_for_diff_k_break_inset_dRgbolsqdt_vs_k_break}). 	

	It would be very interesting to compare the simulation results to experiments on model systems: fully realistic numerical models of digestive processes are currently out of reach, but developing simpler numerical and experimental model systems in tandem will help to identify and understand the most important physical and chemical processes which contribute to digestion. This will help to address the long-term goal of designing healthier foods.	

\section*{Conflicts of interest}

	There are no conflicts to declare.

\section*{Acknowledgements}
	J.K.N.'s work was funded through EU's Horizon 2020 Program, Grant No. 665440 (ABIOMATER). 
	A.D. was supported by a Royal Commission for the Exhibition of 1851 Research Fellowship and by the Novo Nordisk Foundation (Grant Agreement No. NNF18SA0035142). 
	A.Z. acknowledges funding from the Austrian Science Fund (FWF) through a Lise-Meitner Fellowship (Grant No. M 2458-N36).
	J.M.Y. acknowledges funding from the BBSRC (Grant No. BB/P02386X/1).
	We thank Prof. Serafim Bakalis for suggesting this problem to us. 
	We thank Profs. Christos Likos, Knut Drescher and Gary Frost, as well as members of the MMOD consortium, for helpful discussions.
	
\section*{Supplementary Movies}
	
	This is a list of the supplementary movies that illustrate some of the simulations underlying the data presented in the paper.
	\begin{enumerate}
		
		\item An aggregate with no cross-linking in a quiescent fluid.
		
		\item An aggregate stabilized by physical cross-links in a quiescent fluid.
		
		\item An aggregate dispersed through chemical breakdown of its constituent molecules.
		
		\item An aggregate tumbling/tank-treading in moderately strong shear.
		
		\item An aggregate under the same conditions as in 4. from which a single polymeric chain breaks off.
		
		\item An aggregate breaking down in strong shear.
		
		\item An aggregate under the combined action of chemical breakdown and shear.
	\end{enumerate}

\printbibliography

\end{document}